\documentstyle[prl,aps]{revtex}

\begin{document}

\draft
\twocolumn

\title{A search for a border between classical and quantum worlds}

\author{Robert Alicki}

\address{Institute of Theoretical Physics and Astrophysics, University
of Gda\'nsk, Wita Stwosza 57, PL 80-952 Gda\'nsk, Poland}

\date{\today}
\maketitle

\begin{abstract}
The effects of environmental decoherence on a mass center position of 
a macroscopic body are studied using the linear quantum Boltzmann equation.
It is shown that under realistic laboratory conditions these effects can be
essentially eliminated for dust particles containing $10^{15}$ atoms.
However, the initial maxwellian velocity distribution puts limits on
interference type measurements restricting a diameter of a particle to
at most 10 nm. The results are illustrated by the analysis of the recent experiments involving $C_{60}$ and $C_{70}$ fullerenes.

\end{abstract}


Despite its long history the problem of transition between macroscopic
and microscopic worlds remains a fundamental issue in the discussion of 
foundations of quantum mechanics [1,2]. The simplest model ilustrating this topic
is a macroscopic body moving in a slowly varying gravitational potential. To describe
its motion we use the position of its mass center ${\bf x}(t)$ as a
collective variable. One expects that for all practical purposes 
 ${\bf x}(t)$ is well-localized and evolves according to the Newton equation
$$
{d^2\over dt^2}{\bf x}(t) = - {\bf {\nabla}} U({\bf x}(t))
\eqno(1)
$$  
while apparently quantum delocalized states corresponding to 
macroscopically extended wave packets $\Psi({\bf x},t)$ satisfying
the Schr\"odinger equation
$$
-i\hbar {\partial\over\partial t}\Psi({\bf x},t) =
-{\hbar^2\over 2 M} \Delta \Psi({\bf x},t) + V({\bf x})\Psi({\bf x},t)
\eqno(2)
$$
do not appear ($V({\bf x}) = MU({\bf x})$).
\par
In the literature there are discussed at least four types of mechanisms
leading to localization phenomena (or wave function collapse) of above.
\par
1) {\it Environmental decoherence.} Quantum coherence is destroyed by
scattering processes with particles of an environment both massive and massles 
(photons)[3]. Emission and absorption of thermal photons must be
also included.
\par
2) {\it Decoherence by Bremsstralung.} Electric charges moving in a slowly varying 
potential decohere by emission of soft photons [4]  
\par
3) {\it Wave function collapse by gravity.} Here the exact mechanism is not
known due to the absence of an ultimate theory of quantum gravity
but several models were proposed [5].
\par
4) {\it Spontaneous localization theories.} Fundamental stochastic
or/and nonlinear modifications of the Schr\"odinger equation are proposed which
are neglible at the atomic scale but become relevant for macroscopic bodies
[6].
\par 
In the following we propose (a more accurate than in [3]) description of the first,
most conventional mechanism, and the only one which is very sensitive to the
temperature and the density of environmental particles. Denoting by 
${\bf X}$ and ${\bf P}$ the operators of mass center and total momentum of the 
body we have for any wave vector ${\bf k}$
$$
e^{i{\bf kX}} {\bf P} e^{-i{\bf kX}} = {\bf P} + \hbar{\bf k}\ .
\eqno(3) 
$$
The effect of a collision with a gas particle and emission, absorption 
or scattering of a photon 
is a transfer of momentum $\hbar {\bf k}$ which changes the total
momentum as described by eq.(3) independently of the detailed microscopic
mechanism of energy redistribution. Assuming the statistical independence of different
momentum transfer events (called simply {\it collisions}) we propose the following quantum linear Boltzmann
equation (a special case of a quantum  Markovian  master equation [7]) describing  
time evolution of the reduced density matrix of the center of mass subsystem
$$
{d\over dt} \rho = - {i\over \hbar} [H , \rho] + L\rho
\eqno(4)
$$
where $H = {1\over 2M}{\bf P}^2 + V({\bf X})$,
$$
L\rho = \int d^3{\bf k}\ n({\bf k})\Bigl( e^{-i{\bf kX}} \rho\ e^{i{\bf kX}} 
-\rho\Bigr)
\eqno(5)
$$
and $n({\bf k})$ is a density of collisions per unit time leading to the momentum
transfer $\hbar{\bf k}$. Assuming rotational invariance i.e. $n({\bf k})= 
n( k), k=|{\bf k}|$ we can introduce the following parametrization
$$
4\pi k^2 n(k) = {\cal N} \nu(k)\ ,\ {\cal N} =
4\pi\int_0^{\infty} dk\ k^2\ n(k)\ .
\eqno(6)
$$
Here $\nu(k)$ is a probability density of collisions and ${\cal N}$ their total number per time unit. 
\par
The generator $L$ in the position representation reads
$$
(L\rho)({\bf x}|{\bf y}) = -\gamma (|{\bf x}-{\bf y}|) \rho({\bf x}|{\bf y})
\eqno(7)
$$
where
$$
\gamma(r) = {\cal N}\int_0^{\infty}dk\ \nu(k)\Bigl( 1- {\sin kr\over kr}\Bigr)\ .
\eqno(8)
$$ 
Introducing the average wave vector ${\bar k}$ defined by
$$
{\bar k}^2 = \int_0^{\infty}dk\ k^2\ \nu(k) 
\eqno(9)
$$ 
we obtain simple formulas for the decay rates of the off-diagonal matrix elements $\rho({\bf x}|{\bf y})$ in two regimes:
\par 
for  $|{\bf x}-{\bf y}| >> {\bar {\lambda}} = 2\pi{\bar k}^{-1}$
$$
\gamma(|{\bf x}-{\bf y}|)\simeq {\cal N} 
\eqno(10)
$$ 
\par 
for  $|{\bf x}-{\bf y}| << {\bar {\lambda}}$
$$
\gamma(|{\bf x}-{\bf y}|)\simeq {\cal N}{\bar k}^2\ |{\bf x}-{\bf y}|^2\ .  
\eqno(11)
$$ 
In order to analyse an experiment which takes time $t$ between the preparation of a quantum state and its measurement it is convenient to introduce two new parameters:
$$
{\rm coherence\ factor}\ \ \ \Gamma = e^{-t{\cal N}}
\eqno(12)
$$
and 
$$
{\rm coherence\ length}\ \ \ l_{coh} = {{\bar {\lambda}}\over 2\pi\sqrt{t{\cal N}}}
\eqno(13)
$$ 
which, according to eq.(11) gives the maximal distance $|{\bf x}-{\bf y}|$ such that the corresponding off-diagonal elements decohere less than by a factor $e^{-1}$. The parameter
$l_{coh}$ puts an upper bound on the dimensions of diffraction grating 
(slit width and period) which can produce interference patterns.
\par 
To estimate the effect of thermal photons absorption (or emission) for
a macroscopic body of a radius $R$ treated as a black body we calculate
${\cal N}$ as a number of photons with Planck density entering
a surface of a ball per unit time
$$
{\cal N} = {1\over 4} (4\pi R^2 c){1\over \pi^2}
\int_0^{\infty}{k^2\ dk\over e^{c\hbar k/k_BT}-1}
\simeq 0.8\ R^2 c \Bigl({k_BT\over \hbar c}\Bigr)^3\ .
\eqno(14)
$$
The corresponding decoherence time $\tau = {\cal N}^{-1}$ is given by
$$
\tau [sec]\approx  10^{-17} ( R[m])^{-2}(T[K])^{-3}\ .
\eqno(15)
$$
It follows from eq.(15) that the $3K$ background radiation alone washes out all quantum coherence effects described by eq.(2) for the macroscopic bodies 
with $R>> {\bar{\lambda}}= 2\pi\hbar c/k_BT\approx 10^{-3}m$.  
\par
Consider now a laboratory experiment performed at temperatures of the order of $T\approx 1K$, high vacuum of $n_0\approx 10^9 particles/m^3$ (mass of the gas 
particle $m\approx 10^{-25}kg$) and with a "small macroscopic"
body , say a metallic ball of a radius $a= 10^{-5}m$ containing $\approx 10^{15}$ 
atoms. Because at low temperatures the metallic body is almost a perfect
conductor and its radius is much smaller than the thermal radiation
wavelength the leading decoherence factors are the scattering of
a low density gas particles and the Rayleigh scattering of thermal photons.  A number of collisions per unit time for the former is given in terms of the
average thermal velocity $v_{th} = \sqrt{8k_BT/\pi m}$
$$
{\cal N}_{gas} = {1\over 2}(4\pi a^2\ v_{th}) n_0 =
4\sqrt{2\pi}\ a^2\ n_0\ \sqrt{k_BT/m}\ .
\eqno(16)
$$
The Rayleigh scattering is characterized by the $k$-dependent cross-section [8]
$$
\sigma (k) = {10\pi\over 3}k^4a^6\ .
\eqno(17)
$$      
and leads to 
$$
{\cal N}_R = {1\over 2} \Bigl({10\pi\over 3} a^6 c\Bigr){1\over \pi^2}
\int_0^{\infty}{k^6\ dk\over e^{c\hbar k/k_BT}-1}
$$
$$
\simeq 380\ c\ a^6\Bigl({k_BT\over \hbar c}\Bigr)^7
\eqno(18)
$$
A straightforward calculation with the parameters of above yields
$$
{\cal N}_{gas} \approx 20 [sec^{-1}]\ , {\cal N}_R\approx 500[sec^{-1}]\ .
\eqno(19)
$$ 
Both contributions are comparable in this regime and display quite different
temperature and  radius dependence. Hence in principle the onset of environmental decoherence might be observed and well separated from the 
other hypothetical mechanisms like {\it gravitational collapse} and 
{\it spontaneous localization} which incidentally are supposed to be of the comparable magnitude for a body containing 
$10^{15}$ atoms [5][6](the mass center motion of an electrically neutral body should not produce {\it Bremsstralung}).
\par
Unfortunately, the main obstacle is now a possibility of preparing and detecting quantum delocalised states. Any interference type experiment
demands that the de Broglie wavelength $\Lambda = 2\pi\hbar / MV$ is comparable with the width of the slits $d$. In all existing experiments
starting from the historical Young one till the recent ones performed by Zeilinger group [9] the ratio $\delta = \Lambda/d$ is between
$10^{-4} -1$. Obviously, for diffraction gratings we have a geometrical condition 
$$
\Lambda = \delta d\geq 2 \delta a\ .
\eqno(20)
$$
On the other hand $V$ cannot be smaller than the thermal 
velocity $V_{th}=\sqrt{8k_BT/\pi M}$ what gives
$$
\Lambda\leq {2\pi\hbar\over M V_{th}} = {(\pi)^{3/2}\hbar\over \sqrt{(2Mk_BT)}}\ .
\eqno(21)
$$
Puting $M = (4/3)\pi a^3 \kappa$ where the density of the body $\kappa
\approx 10^4 kg/m^3$ and a rather optimistic value for $\delta = 10^{-5}$ we obtain from eqs.(20,21) the final condition for the successful interference type experiment
$$
a \leq \delta^{-2/5} \bigl(\hbar^2/\kappa k_B T\bigr)^{1/5} 
\approx 10(T[K])^{-1/5}[nm] \ .
\eqno(22)
$$
The very weak temperature dependence of the right hand side of eq.(22)
makes rather impossible to go essentially far beyond the $nm$ scale with
traditional interference measurements. 
\par
The recent successful experiments involving $C_{60},C_{70}$ [9], the molecules 
with $a\simeq 0.5 nm$, lie not far from the border between classical
and quantum worlds established by eq.(22).The authors rightfully argued that
decoherence effects can be neglible under the conditions of their
experiments. We can quite precisely estimate the decoherence magnitude
using their data. First we have to compute the total number of 
collisions $t{\cal N}$ during the time of flight $t$ of a fullerene due to emission, 
absorption and Rayleigh scattering of radiation and scattering
of gas particles. For the emission the authors estimate
$t{\cal N}_1\sim 3.5$. As the environment temperature $T_1 \approx 300 K$
is much lower than the temperature of the fullerene molecule $T_2\approx
900 K$ then due to eq.(14) absorption can be neglected. The same holds due 
to eq.(18) for the
Rayleigh scattering because  the radius $a$ is much smaller than the average
radiation wavelength $\lambda_{T_1}\approx 10\mu m$. The number of collisions 
with gas particles is estimated to be $t{\cal N}_2\approx 10^{-2}$ and can
be neglected also.  As  ${\bar {\lambda}}\approx 10\mu m $ the coherence length (13)
 $l_{coh}\approx 1\mu m$ -- the value which is still essentially
larger than the width of the slits (50nm) and their separation (100nm).
It follows that the diffraction picture is not destroyed by decoherence which 
reduces the effective collimation width only.

\par
It was shown that under realistic laboratory conditions the environmental 
decoherence of the center of mass position can be eliminated on the time
scale of {\it miliseconds} for macroscopic dust particles containing $10^{15}$ 
atoms.
Nevertheless, the emerging quantum coherence effects are completely 
overshadowed by the initial maxwellian distribution of the dust particle velocity 
at least in all standard diffraction - interference type experiments. Therefore, 
only completely new ideas concerning preparation and measurement
of spatially extended quantum states might push the border between
quantum and classical worlds beyond the scale of {\it nanometers}. 
On the other hand the experiments involving large molecules of a diameter
less than $10nm$ are feasible an can provide interesting information concerning 
the detailed mechanism of environmental decoherence.

\par

\acknowledgments
The author thanks M. \.Zukowski and M. and R. Horodecki for discussions.
The work is supported by the Grant KBN 2PO3B 04216 .

\end{document}